\documentclass[aps,prd,reprint,longbibliography,nofootinbib,superscriptaddress,floatfix]{revtex4-2}
\usepackage[T1]{fontenc}
\usepackage{lmodern}
\usepackage[utf8]{inputenc}
\usepackage{amsmath,amssymb,bm}
\usepackage{graphicx}
\usepackage{xcolor}
\definecolor{myblue}{RGB}{71,81,200}
\usepackage{hyperref}
\hypersetup{colorlinks=true,linkcolor=myblue,citecolor=myblue,urlcolor=myblue}
\newcommand{\Rew}{\operatorname{Re}}
\newcommand{\Imw}{\operatorname{Im}}
\newcommand{\sech}{\operatorname{sech}}
\newcommand{\dd}{\mathrm{d}}

\AtBeginDocument{}

\begin{document}

\title{Spectral instability of charged scalar quasinormal modes and superradiance: Stabilizing the unstable}

\author{Kyriakos Destounis}

\affiliation{CENTRA, Departamento de F\'isica, Instituto Superior T\'ecnico -- IST, Universidade de Lisboa -- UL, Avenida Rovisco Pais 1, 1049-001 Lisboa, Portugal}

\affiliation{School of Applied Mathematical and Physical Sciences -- SEMFE, Physics Department, National Technical University of Athens -- Polytechnic Campus, Zografou 15772, Athens, Greece}

\begin{abstract}
The spectral instability of black holes has become a central topic of the black-hole spectroscopy program. Small changes to the black-hole scattering potential are known to produce disproportionately large migrations of its quasinormal modes in the complex plane, but it is unclear whether this sensitivity carries over to energy extraction through superradiance. We study this relation for massless charged scalar fields on Reissner--Nordstr\"om and Reissner--Nordstr\"om--de Sitter backgrounds. A localized positive potential bump, small in mass-scaled height, is displaced outward from the light ring while the original radiative boundary conditions are preserved. In the asymptotically flat case, a quasinormal branch enters the superradiant frequency regime while remaining damped. By searching for competing modes, we show that the least-damped positive-frequency mode also enters this regime, after a series of mode overtakings driven by quasinormal-mode migration. The real-frequency amplification, nevertheless, is only weakly affected and develops no pronounced narrow resonance. In de Sitter black holes, the bump can move an initially growing monopole mode into the damped half of the complex plane, stabilizing selected configurations that are initially modally unstable. After stabilization, the peak superradiant amplification can remain above both its unperturbed value and a gain of $100\%$; it falls below $100\%$ only for more distant bump locations. A prescribed positive perturbation can thus suppress modal growth without changing the superradiant frequency bounds or immediately suppressing the scattering peak.
\end{abstract}
\maketitle

\section{Introduction}\label{sec:introduction}

The observation of gravitational waves from merging black holes (BHs) has made their response to perturbations a central subject of gravitational physics~\cite{LIGOScientific:2016aoc,VIRGO:2014yos,KAGRA:2018plz}. During ringdown, a perturbed black hole (BH) is described, over appropriate time intervals, by a superposition of damped oscillations with complex quasinormal-mode (QNM) frequencies~\cite{Kokkotas:1999bd,Berti:2009kk,Konoplya:2011qq}. Measuring these frequencies and comparing them with the predictions of general relativity underlies BH spectroscopy~\cite{Echeverria:1989hg,Dreyer:2003bv,Berti:2005ys,Giesler:2019uxc,Isi:2019aib,Destounis:2023ruj,Berti:2025hly}. The interpretation of such measurements requires understanding both the spectrum of an idealized BH and its response to changes in the surrounding scattering potential, that can pose as a toy model of environmental effects \cite{Barausse:2014tra,Cheung:2021bol,Berti:2022xfj,Li:2024npg,Courty:2023rxk,Destounis:2025dck,Cardoso:2021wlq,Cardoso:2022whc,Destounis:2022obl,Destounis:2025tjn,Kouniatalis:2025itj,Pezzella:2024tkf,Xu:2026qrg}.

QNM boundary conditions allow radiation to escape through the event horizon and the outer boundary. The resulting spectral problem is non-self-adjoint, and its eigenvalues can be sensitive to small perturbations, i.e., they exhibit a spectral instability \cite{Jaramillo:2020tuu,Jaramillo:2021tmt,Yang:2024vor,Cardoso:2024mrw,Ianniccari:2024ysv,Torres:2023nqg,Boyanov:2024fgc,Warnick:2024usx,Warnick:2013hba,May:2026qpf}. Early examples of this sensitivity~\cite{Nollert:1996rf} have been placed in a broader framework using the notion of the pseudospectrum~\cite{Jaramillo:2020tuu,Jaramillo:2022kuv,Destounis:2021lum,Destounis:2023nmb,Boyanov:2022ark,Arean:2024afl,Arean:2023ejh,Cownden:2023dam,Garcia-Farina:2024pdd,Gasperin:2021kfv,Boyanov:2023qqf,Sarkar:2023rhp,Luo:2024dxl,Cai:2025irl,Destounis:2026qne,Trefethen_Embree_book}. Studies of charged and cosmological BHs backgrounds likewise reveal spectral instability~\cite{Destounis:2021lum,Sarkar:2023rhp,Destounis:2023nmb,Courty:2023rxk}. A direct diagnostic consists of adding a small localized bump to the effective potential and moving it away from the main scattering region. Even the least-damped mode can change substantially: a continued branch can migrate, other branches can overtake it, and resonances associated with the enlarged scattering cavity can become relevant~\cite{Cheung:2021bol}. Following one pole is, thus, insufficient to establish the behavior of the least-damped part of the spectrum.

Large changes in individual poles need not translate into equally large changes in real-frequency observables. Studies of greybody factors find substantial robustness under small potential deformations~\cite{Rosato:2024arw}, while also identifying sensitivity at higher frequencies and for sharper perturbations~\cite{Oshita:2024fzf}. A complementary complex-angular-momentum analysis shows that a perturbed Regge-pole spectrum can still reconstruct stable scattering quantities~\cite{Torres:2023nqg}. These results motivate comparing spectral migration with the real-frequency response directly, rather than inferring one from the other. Here, a charged test scalar field provides a tractable model with an amplifying scattering channel; it is not a calculation of a gravitational-wave signal.

Superradiance \cite{Brito:2015oca} provides a complementary probe of a similar boundary-value problem. An incident scattering bosonic wave can extract rotational or electromagnetic energy from a BH, producing a reflected flux larger than the incident flux~\cite{Bekenstein:1973mi}. For a charged scalar on a charged, nonrotating background, the sign of the horizon flux is controlled by the frequency measured relative to the horizon electrostatic potential. This fixes a real-frequency regime in which superradiant amplification is possible. The existence of that regime does not by itself imply a growing solution of the homogeneous perturbation equations. Modal instability requires a QNM in the upper half of the complex-frequency plane, whereas amplification is a property of a real-frequency scattering solution.

The asymptotically flat Reissner--Nordstr\"om (RN) geometry is a useful reference system. Charged massless scalar waves can be superradiantly amplified, while the unperturbed modes used in the present calculations are damped and lie outside the positive-frequency superradiant regime~\cite{Richartz2014,Benone:2015bst,Mollicone:2024lxy}. Introducing a positive cosmological constant changes the problem qualitatively. Reissner--Nordstr\"om--de Sitter (RNdS) BHs admit a charged-scalar monopole instability in parts of their parameter space, despite imposing an outgoing condition at the cosmological horizon~\cite{Zhu:2014sya,Konoplya:2014lha,Cardoso:2018nvb,Dias:2018ufh,Destounis:2019hca}. The unstable branch can be connected to the neutral, static monopole and to the de Sitter family~\cite{Destounis:2019hca}. The cosmological horizon is not a reflecting wall; the instability is a property of the full charged-field boundary-value problem.

Long-lived poles near the real axis can produce strong resonant amplification in RNdS, and such peaks already occur in both stable and unstable configurations~\cite{Mascher:2022pku,Destounis:2022rpk}. Their height and shape depend on the coupling of the pole to the scattering channels and on the non-resonant response, in addition to its frequency. A pole's real part entering the superradiant regime is, hence, insufficient to predict a narrow or large amplification peak. The question addressed here is how these diagnostics change along a controlled deformation of a given system. We vary the position of a localized positive bump while holding the geometry and charge coupling $qQ$ fixed. The superradiant bounds, consequently, remain unchanged, allowing a change in modal stability to be separated from a change in the kinematically allowed energy-extraction regime.

We ask, first, whether RN migration into that regime survives when competing modes and changes of damping order are included, and, second, whether a positive perturbation can suppress an existing RNdS instability and how the amplification then changes along the same displacement sequence. We use a localized P\"oschl--Teller profile added to the scalar radial potential, with a horizon window that preserves the boundary conditions. This is a phenomenological modification of propagation on a fixed background. Self-consistent environmental geometries provide a complementary approach~\cite{Cardoso:2021wlq,Stelea:2023yqo,Mollicone:2024lxy}. Our profile is specified in areal radius; its relation to the tortoise-radius displacement experiment of Ref.~\cite{Cheung:2021bol} is discussed when defining the model.

The results exhibit two distinct behaviors. In RN, large fractional frequency changes and overtaking can place the least-damped positive-frequency mode found inside the superradiant regime, while the resolved modes remain damped and the sampled scattering curves show no pronounced narrow resonance. In RNdS, a positive bump can move the initially unstable monopole mode across the real axis, and independent root counts support the absence of growing monopole modes at selected target configurations. Amplification remains enhanced after this transition and decreases only at larger displacement. The contribution is, thus, the connection between pole migration, mode ordering, and scattering along a fixed-geometry deformation, including the separation between stabilization and suppression of the amplification peak. We do not compute a charged-field pseudospectrum or a time-domain waveform.

We use geometrized units $G=c=1$, the time convention $e^{-i\omega t}$, and the gauge $A_t=-Q/r$. Numerical results are expressed in units of the BH mass $M$. In particular, the cosmological constant and bump amplitude are reported as $\Lambda M^2$ and $\epsilon M^2$.

\section{Background geometry and perturbation model}\label{sec:geometry}
\subsection{Geometry and coordinates}

The RN and RNdS metrics are static and spherical symmetric solutions to electrovacuum Einstein field equations, where the former is asymptotically flat ($\Lambda=0$), while the latter is asymptotically de Sitter ($\Lambda>0$). Their line element has the form
\begin{align}\label{eq:metric}
 d s^2&=-f(r)d t^2+f(r)^{-1}d r^2+r^2d\Omega^2,
\end{align}
where $d\Omega^2=d\theta^2+\sin^2\theta\,d\phi^2$ is the metric of the unit two-sphere, and the lapse functions for RN and RNdS are
\begin{align}
f_{\rm RN}(r)&=1-\frac{2M}{r}+\frac{Q^2}{r^2},\qquad\quad\,\qquad\text{(RN)},\\
    f_{\rm RNdS}(r)&=1-\frac{2M}{r}+\frac{Q^2}{r^2}-\frac{\Lambda r^2}{3},\qquad\text{(RNdS)}.
\end{align}
The electromagnetic potential is $A_\mu=(-Q/r,0,0,0)$. A subextremal RN geometry possesses an inner Cauchy horizon at $r=r_-$, and an event horizon at $r=r_+$, while a subextremal RNdS geometry possesses the two aforementioned horizons and a cosmological horizon at $r=r_c$. The three horizons satisfy $r_-<r_+<r_c$, which are given by the positive roots of $f(r)$. We solve the perturbation problem in the static region $r_+<r<r_c$. For RN, the outer boundary is spatial infinity and $r_\pm=M\pm\sqrt{M^2-Q^2}$.

The metric and horizon locations are unchanged when the local weak perturbation is introduced. In particular, varying the field coupling $qQ$ or the bump amplitude does not move either horizon. The radius of the photon sphere location where circular unstable null orbits reside is (for RN and RNdS)
\begin{equation}\label{eq:photon}
 r_{\rm ph}=\frac{3M+\sqrt{9M^2-8Q^2}}{2}.
\end{equation}
This supplies a geometric reference for the bump's position; the peak of the monopole scalar potential need not coincide with it. Throughout, $a$ denotes the bump center in areal radius.

\subsection{Charged scalar equation}\label{sec:field}

A massless scalar of charge $q$ satisfies the charged Klein-Gordon equation
\begin{equation}\label{eq:KG}
 D^\mu D_\mu\psi=0,\qquad D_\mu=\nabla_\mu-iqA_\mu.
\end{equation}
Under the assumption of an ansatz $\psi=r^{-1}\Psi_{\ell m}(r)Y_{\ell m}(\theta,\phi)e^{-i\omega t}$, with a Fourier decomposition that renders growing modes to satisfy $\textrm{Im}(\omega)>0$, and the definition of the tortoise coordinate $d r_*/d r=f^{-1}$, the the unperturbed (bumpless) radial equation becomes
\begin{equation}\label{eq:radial0}
 \frac{\dd^2\Psi}{\dd r_*^2}
 +\bigl[(\omega-\Phi)^2-U_\ell\bigr]\Psi=0.
\end{equation}
Here,
\begin{equation}\label{eq:potential}
 \Phi(r)=\frac{qQ}{r},\qquad
 U_\ell(r)=f(r)\left[\frac{\ell(\ell+1)}{r^2}+\frac{f'(r)}{r}\right],
\end{equation}
where a prime denotes $d/d r$. Equivalently, the coefficient in brackets is $\omega^2-2\omega\Phi-V_\ell$, with $V_\ell=U_\ell-\Phi^2$. Keeping the electrostatic term inside $(\omega-\Phi)^2$ makes the horizon frequency shifts explicit. A negative region of $V_\ell$ alone is not a stability criterion for this charged, frequency-dependent eigenvalue problem.

\subsection{Localized perturbation and horizon taper}

We prescribe the bump modification to the BH potential as
\begin{align}\label{eq:bump}
 U_\ell&\rightarrow U_\ell+\delta V,\nonumber\\
 \delta V(r)&=\epsilon\,\sech^2\!\left(\frac{r-a}{w}\right)W(r),
 \qquad w=M,
\end{align}
and solve the new perturbed eigenvalue problem
\begin{equation}\label{eq:radial}
 \frac{\dd^2\Psi}{\dd r_*^2}
 +\bigl[(\omega-\Phi)^2-U_\ell-\delta V\bigr]\Psi=0.
\end{equation}
The amplitude $\epsilon>0$ has dimensions $M^{-2}$. This modification is applied after separation of the test-field equation. We note here that no corresponding stress-energy tensor or backreacted metric is assumed. Nevertheless, there are techniques that can be used to construct a spacetime with matter that leads to the aforementioned potential (see \cite{Cheung:2021bol,Courty:2023rxk} appendices, though see also \cite{DellaRocca:2026zym}).

The window $W$ is part of the model. Let $d=M/4$ and define the $C^2$ function
\begin{equation}\label{eq:window}
 S(x)=\begin{cases}
 0,&x\leq1,\\
 6t^5-15t^4+10t^3,\quad t=x-1,&1<x<2,\\
 1,&x\geq2.
 \end{cases}
\end{equation}
Then the window will be
\begin{equation}\label{eq:W}
 W(r)=\begin{cases}
 S((r-r_+)/d),&\Lambda=0,\\
 S((r-r_+)/d)S((r_c-r)/d),&\Lambda>0.
 \end{cases}
\end{equation}
Thus, the bump vanishes identically within $0.25M$ of each horizon and retains its full profile beyond $0.5M$. This matters because an untapered areal-radius $\sech^2$ profile is nonzero at a finite horizon, however small its tail, and would change the asymptotic radial equation. The window preserves the usual ingoing and outgoing horizon conditions exactly. Its independence from the reported observables is a separate question, discussed in Sec.~\ref{sec:limitations}.

\subsection{Perturbation scale and coordinate width}\label{sec:bump_scale}

The small parameter $\epsilon M^2$ measures the bump height relative to the BH mass scale. It does not imply a uniformly small fractional change of the local background potential. At centers outside the taper regions, $\delta V(a)=\epsilon$, while $U_0(a)$ becomes small towards the outer
boundary. For example, when $Q/M=0.6$, $\Lambda M^2=10^{-3}$, and $\epsilon M^2=10^{-4}$, the ratios $\delta V(a)/|U_0(a)|$ at $a/M=30,45,48$ are approximately $0.27$, $0.55$, and $0.81$, respectively. These are comparisons of local potential heights, not estimates in a specified operator norm.

Locally expanding the tortoise map around the bump center gives
\begin{equation}\label{eq:bumpwidth}
 \begin{aligned}
 r_*-r_*(a)&=\frac{r-a}{f(a)}-\frac{f'(a)}{2f(a)^2}(r-a)^2+\cdots,\\
 w_*&\simeq\frac{w}{f(a)}.
 \end{aligned}
\end{equation}
If $f$ varies slowly across the bump (in particular,
$w|f'(a)|/f(a)\ll1$ and $w^2|f''(a)|/f(a)\ll1$) and the horizon windows
do not appreciably truncate the profile, its integrated strength is approximately
\begin{equation}\label{eq:bumparea}
 \int\delta V\,d r_*
 =\int_{\text{exterior}}\frac{\delta V(r)}{f(r)}\,d r
 \simeq\frac{2\epsilon w}{f(a)}.
\end{equation}
Thus, displacement changes both the location and the tortoise-coordinate extent of this areal-radius profile. It is not a rigid translation in $r_*$. A fixed-tortoise-width bump defines a different family of perturbations.

For the same $\Lambda M^2=10^{-3}$ examples, the local estimates are $w_*/M\simeq1.58,3.56,5.25$ at $a/M=30,45,48$. There $w|f'(a)|/f(a)\simeq0.03,0.10,0.16$, so the local expansion becomes only moderately accurate at the outermost positions. In RN Case (2), introduced in Table \ref{tab:RN}, the center value at $a/M=100$ is
$\delta V(a)/|U_0(a)|\simeq51.2$, despite $\epsilon M^2=10^{-4}$. Thus ``small'' throughout refers to the mass-scaled height, not to an everywhere small fractional deformation. The tortoise-radius bump used in
Ref.~\cite{Cheung:2021bol} motivates the overtaking analysis, but its numerical transition distances cannot be transferred directly to this areal-radius prescription.

\section{Superradiant scattering}
\label{sec:scattering}

For real $\omega$, define $k_h=\omega-qQ/r_+$ and $k_c=\omega-qQ/r_c$. An incident wave from the cosmological horizon obeys
\begin{equation}\label{eq:scatds}
 \Psi\sim\begin{cases}
 \mathcal{T}e^{-ik_h r_*},&r\to r_+,\\
 \mathcal{I}e^{-ik_c r_*}+\mathcal{R}e^{ik_c r_*},&r\to r_c.
 \end{cases}
\end{equation}
We restrict to $k_c>0$, so the incident channel has positive frequency relative to the cosmological-horizon potential. Its zero-frequency endpoint is excluded from the scattering normalization.

For RN, the outgoing and incoming waves contain a Coulomb logarithmic phase. At large $r$, a suitable phase is
\begin{align}\label{eq:coulomb}
 \Theta(r)&=\omega r_* -qQ\ln(r/M)\nonumber\\
 &=\omega r+(2M\omega-qQ)\ln(r/M)+O(1).
\end{align}
The outer condition becomes $\Psi\sim\mathcal{I}e^{-i\Theta}+\mathcal{R}e^{i\Theta}$, supplemented by inverse-radius corrections in the numerical boundary series. Omitting this logarithmic phase would not give the correct charged-field asymptotics. The event-horizon condition remains that in Eq.~\eqref{eq:scatds}.

Since the coefficients of Eq.~\eqref{eq:radial} are real for real $\omega$, the current
\begin{equation}
 J=\frac{1}{2i}\left(\Psi^*\frac{d\Psi}{d r_*}
 -\Psi\frac{d\Psi^*}{d r_*}\right)
\end{equation}
is conserved. Matching its values at the boundaries gives
\begin{equation}\label{eq:flux}
 |\mathcal{R}|^2=|\mathcal{I}|^2-\frac{k_h}{k_o}|\mathcal{T}|^2,
 \qquad k_o=\begin{cases}\omega,&\mathrm{RN},\\k_c,&\mathrm{RNdS}.\end{cases}
\end{equation}
For positive $qQ$, the superradiant regimes are then
\begin{align}\label{eq:SR}
 0<\omega<\frac{qQ}{r_+}&\quad\mathrm{(RN)},\nonumber\\
 \frac{qQ}{r_c}<\omega<\frac{qQ}{r_+}&\quad\mathrm{(RNdS)}.
\end{align}
The real bump does not change these bounds, since it changes neither the geometry nor the asymptotic electrostatic potentials. We define the amplification gain by
\begin{equation}\label{eq:gain}
 Z_\ell=\left|\frac{\mathcal{R}}{\mathcal{I}}\right|^2-1
 =-\frac{k_h}{k_o}\left|\frac{\mathcal{T}}{\mathcal{I}}\right|^2.
\end{equation}
We write $Z_\ell^{(\epsilon)}$ and $Z_\ell^{(0)}$ to distinguish the perturbed and unperturbed systems. The subscript in $Z_0$ refers only to $\ell=0$. Our percentage plots show $100Z_\ell$. Thus zero denotes no gain, and $100\%$ denotes reflected flux equal to twice the incident flux. A $100\%$ reference line is a useful measure of the magnitude of amplification, but has no role in the modal-stability criterion. Likewise, $100\Delta Z_\ell=100[Z_\ell^{(\epsilon)}-Z_\ell^{(0)}]$ is a difference in percentage points, not a relative percentage change.

The conserved quantity $J$ is the radial wave current. At the boundaries, $J_h=-k_h|\mathcal T|^2$ and $J_o=k_o(|\mathcal R|^2-|\mathcal I|^2)$. Incident and reflected waves are compared at the same outer boundary,
where their frequency factor $k_o>0$ is identical. Their energy-flux ratio is
\begin{equation}\label{eq:energy_ratio}
 \frac{\mathcal F_{\rm ref}}{|\mathcal F_{\rm inc}|}
 =\frac{k_o^2|\mathcal R|^2}{k_o^2|\mathcal I|^2}
 =\left|\frac{\mathcal R}{\mathcal I}\right|^2.
\end{equation}
Here, the energy reference is chosen to make the electrostatic potential vanish at the outer boundary, while common normalization factors cancel out. This establishes the energy-gain interpretation of $Z_\ell$, without identifying $J$ itself with the local stress-energy flux at every radius in the external electric field. Under a constant gauge shift,
\begin{equation}\label{eq:gauge}
 A_t\to A_t+C,\qquad \omega\to\omega-qC,\qquad \Phi\to\Phi-qC.
\end{equation}
Thus $\omega-\Phi$, $k_h$, $k_o$, and the flux condition are gauge invariant.
All frequencies and positive-frequency rankings quoted here use the stated
gauge $A_t=-Q/r$.

For a background supporting a growing QNM, the real-frequency scattering solution is still a well-defined boundary-value response away from a real-axis pole. It is not generally the late-time limit of an initial-value problem: generic initial data also excite growing homogeneous solutions. All comparisons involving unstable configurations below have this frequency-domain meaning.

\section{Quasinormal modes}\label{sec:QNMs}

QNMs satisfy the homogeneous radiative conditions obtained by setting $\mathcal{I}=0$ in the scattering problem. They are ingoing at $r_+$ and outgoing at $r_c$, or outgoing with the Coulomb phase~\eqref{eq:coulomb} at RN infinity. Writing $\omega=\omega_R+i\omega_I$, our convention gives decay for $\omega_I<0$ and exponential growth for $\omega_I>0$. The charged equation and its boundary conditions obey the symmetry
\begin{equation}\label{eq:symmetry}
 (\omega,qQ)\rightarrow(-\omega^*,-qQ).
\end{equation}
Consequently, positive and negative real frequencies are not redundant at a fixed nonzero coupling. We take $qQ>0$, while avoiding negative-real-part frequencies in the spectral checks.

Three different notions of mode ordering are needed. A continued branch is defined by following one pole as a parameter changes. The positive-frequency pole closest to the real axis is selected independently at each position:
\begin{equation}\label{eq:ranking}
 \omega_{+,\min}(a)\in
 \underset{\omega_j:\,\Rew\omega_j>0}{\operatorname{arg\,min}}
 |\Imw\omega_j(a)|.
\end{equation}
Finally, the largest imaginary part,
\begin{equation}\label{eq:gamma}
 \Gamma(a)=\sup_j\Imw\omega_j(a),
\end{equation}
determines whether a growing QNM is present. Numerically, these selections are made over the modes resolved by the search, with separate counts of growing roots at selected RNdS configurations. In an unstable spectrum, the mode minimizing $|\omega_I|$ need not have the fastest growth. We, therefore, avoid identifying all of these orderings with a single universal ``fundamental mode''.

RNdS spectra contain photon-sphere (PS), de Sitter (dS), and near-extremal families of modes~\cite{Cardoso:2017soq,Cardoso:2018nvb,Destounis:2019hca}. Each family is designated by their limiting behavior, not simply damping order at one parameter value. We identify the branches retained in our family figure by continuation in charge, cosmological constant, and bump amplitude. No family assignment is inferred solely from the charge being far from extremality.

\subsection{Numerical methods}

The calculations use an implementation of two-sided shooting and spectral collocation. Horizon Frobenius expansions supply ingoing or outgoing boundary data in the region where the bump vanishes. Matching the solutions and their derivatives yields a Wronskian determinant whose zeros locate QNMs. RN infinity is treated using its charged asymptotic series and an exterior complex contour. Unperturbed RN seed frequencies are checked with a separate continued-fraction implementation~\cite{Richartz2014}. Independent mode discovery uses piecewise Chebyshev collocation of a quadratic eigenvalue problem, resolving the bump and horizon windows. Candidate modes are compared across resolutions and checked by shooting. The collocation formulation follows the general spectral approach to QNM computation~\cite{Jansen:2017oag}.

For real-frequency scattering, boundary expansions and numerical integration are matched to obtain $\mathcal{I}$, $\mathcal{R}$, and $\mathcal{T}$. Flux conservation provides an independent consistency test. Controls vary the expansion order, horizon launch distances, integration tolerances, matching location, and bump resolution. RN checks also vary the outer contour or scattering boundary. The RN migration curves use a $0.25M$ position step, checked against $0.5M$ steps. RNdS amplification peaks are refined by direct frequency optimization. Appendices~\ref{app:matching} and~\ref{app:spectral} give the boundary factors, matching determinant, and RNdS quadratic spectral formulation. Appendix~\ref{app:counts} derives the growing-mode bounds, states the root-counting prescription, and summarizes the available count results and their numerical limitations.

\section{Spectral migration and changes of mode ordering}\label{sec:migration}

We restrict the results to $\ell=0$. In the following, spectral instability denotes pronounced sensitivity of QNM frequencies or their ordering to a small prescribed perturbation. It does not denote exponential growth. We report the bump amplitude and frequency displacement explicitly, without interpreting their ratio as a pseudospectral bound in an unspecified operator norm.

\subsection{RN: continued branches and overtaking}\label{sec:RNmodes}

The two RN examples are summarized in Table~\ref{tab:RN}. Their different charges, charge couplings, and bump amplitudes prevent assigning the difference between their trajectories to any one parameter alone.
\begin{table}[b]
\caption{RN setups. Both have $\ell=0$, $\Lambda=0$, and $w/M=1$. The last column gives the dimensionless upper superradiant boundary.}
\label{tab:RN}
\begin{ruledtabular}
\begin{tabular}{ccccc}
Case & $Q/M$ & $qQ$ & $\epsilon M^2$ & $MqQ/r_+$\\
(1)&0.99&0.99&$10^{-6}$&0.867609\\
(2)&0.6&0.2&$10^{-4}$&0.111111
\end{tabular}
\end{ruledtabular}
\end{table}

\begin{figure*}[t]
\includegraphics[width=\textwidth]{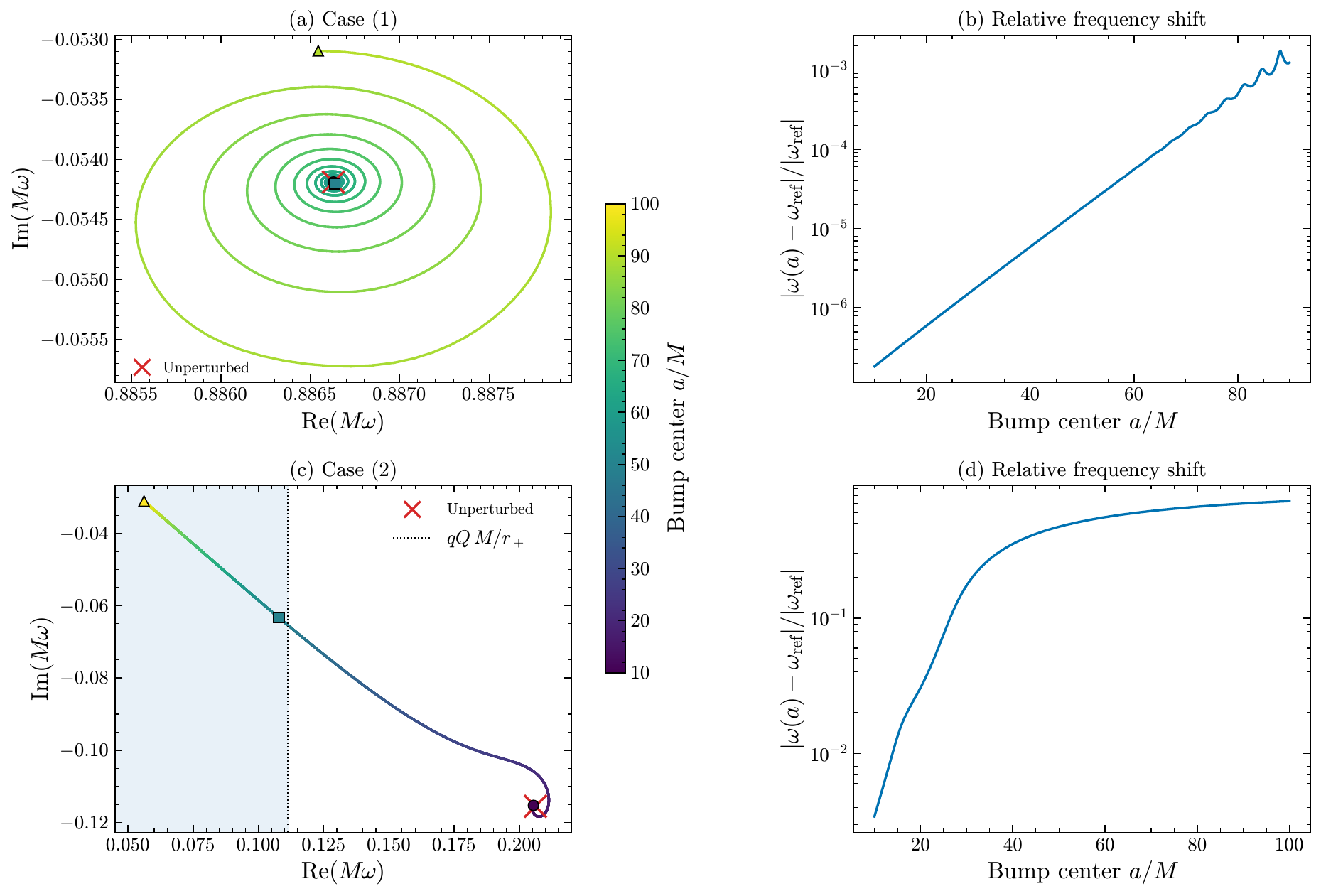}
\caption{Migration of a single continued charged-scalar monopole QNM in RN. The upper and lower rows show Cases (1) and (2) of Table~\ref{tab:RN}, respectively, with bump width $M$. Left: complex-frequency trajectories colored by the areal bump center $a/M$, covering $10\leq a/M\leq90$ in Case (1) and $10\leq a/M\leq100$ in Case (2). The unperturbed mode is marked by a red star, while the circle, square, and triangle mark $a/M=10$, $50$, and the final position, respectively. In Case (1) the first two markers lie close to the unperturbed mode, as the displacement there is below $10^{-4}$. Right: relative displacement $|\omega(a)-\omega_{\rm ref}|/|\omega_{\rm ref}|$ from the corresponding unperturbed frequency $\omega_{\rm ref}$. The displayed superradiant boundary is $\Rew\omega=qQ/r_+$. In Case (2), the tracked branch enters the superradiant regime while remaining damped. A continuous branch need not remain the least-damped mode. The changes of ordering are examined in Fig.~\ref{fig:RNovertaking}.}
\label{fig:RNmigration}
\end{figure*}

We first follow the fundamental mode branch without considering mode overtaking as the bump location is moved awat from the PS. Figure~\ref{fig:RNmigration} demonstrates the trajectories of the fundamental PS modes in the complex plane. The unperturbed seed frequencies are
\begin{align}
 M\omega^{(1)}_{\rm ref}&=0.886628425-0.054188096i,\nonumber\\
 M\omega^{(2)}_{\rm ref}&=0.206130214-0.115534813i.
\end{align}
Both are damped and lie above their respective superradiant upper bounds. In Case (1), the continued pole reaches approximately $0.886543463-0.053095140i$ at $a/M=90$, a relative displacement $|\omega-\omega_{\rm ref}|/|\omega_{\rm ref}|\simeq1.23\times10^{-3}$. Its real part remains outside the superradiant regime. We restrict this continuous trajectory to $90M$ because finer continuation beyond that point exposed a branch-association ambiguity. Independent searches at larger positions are not a validation of the excluded continuation.

Case (2) shows a much larger migration. At $a/M=100$, the continued pole is approximately $M\omega=0.05609818-0.03102951i$, with a relative displacement of $0.729$. A change of this size for $\epsilon M^2=10^{-4}$ demonstrates strong spectral sensitivity to the prescribed potential bump perturbation. The first sampled position at which this branch satisfies the superradiant relation $0<\omega_R<qQ/r_+$ is $a/M=48.5$. Its imaginary part remains negative throughout the displayed trajectory.

\begin{figure*}[t]
\includegraphics[width=\textwidth]{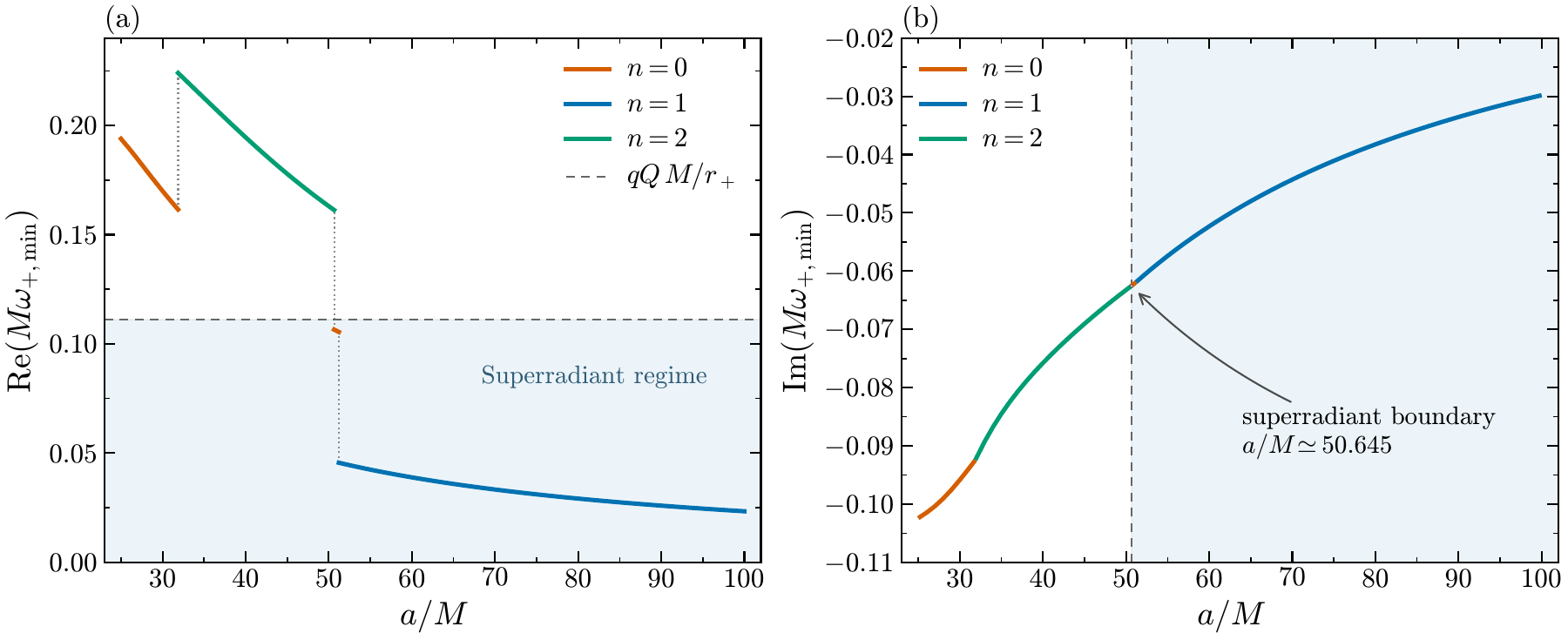}
\caption{Real (left) and imaginary (right) parts of the least-damped positive-frequency QNM found in RN Case (2): $Q/M=0.6$, $qQ=0.2$, $\ell=0$, $\epsilon M^2=10^{-4}$, and $w/M=1$. The dense competing-branch continuation covers $25\leq a/M\leq100$. Labels $n=0,1,2$ are the fixed branch identifiers defined in the text. The selected branch changes at $a/M\simeq31.874$, $50.645$, and $51.213$. Dotted vertical connections in the real-frequency panel indicate changes of the selected branch. The selected positive-frequency mode first enters the superradiant regime through the overtaking near $a/M=50.645$. The shaded portion of the imaginary-frequency panel marks positions satisfying the real-frequency superradiant condition, though it does not indicate modal growth. All displayed modes are damped. Less-damped negative-frequency QNMs are excluded from this positive-sector selection.}
\label{fig:RNovertaking}
\end{figure*}

The next question is whether entering the superradiant regime persists after overtaking of modes takes place. An overtaking instability occurs when the initial fundamental QNM is overtaken by the first overtone at specified bump locations. Figure~\ref{fig:RNovertaking} addresses this using six continued positive-frequency QNMs that do not have initially the smallest in absolute value imaginary part. The labels $n=0,1,2$ in this figure denote fixed branches ordered by damping at $a/M=25$. They are not established unperturbed RN overtone numbers. The selected positive-frequency mode changes in the sequence
\begin{equation}\label{eq:switches}
 n=0\ \rightarrow\ n=2\ \rightarrow\ n=0
 \ \rightarrow\ n=1,
\end{equation}
at approximately $a/M=31.874$, $50.645$, and $51.213$. The return to $n=0$ reflects an unconventional crossing of damping rates with respect to Ref. \cite{Cheung:2021bol}.

The transition at $a/M\simeq50.645$ is particularly informative. The selected real part jumps from about $0.16129/M$ to $0.10656/M$, while the two imaginary parts coincide at approximately $-0.06255/M$. Because the upper superradiant boundary is $0.11111/M$, the selected positive-frequency mode enters the regime by a change of ordering in the perturbed fundamental mode. The originally continued branch had crossed this boundary earlier, when it was not the selected positive-frequency mode.

At $a/M=100$, the selected positive-frequency pole is approximately
\begin{equation}\label{eq:RNwinner}
 M\omega_{+,\min}=0.02337449-0.02980694i.
\end{equation}
The earlier single-branch pole remains present but is more strongly damped. A negative-frequency mode, $M\omega\simeq-0.01252429-0.02402712i$, is less damped than either positive pole. The statement demonstrated here is therefore about the least-damped \emph{positive-frequency} mode found, not the globally least-damped pole across both signs. These resonances are not quasibound states. Their pattern is compatible with an enlarged scattering cavity, but the frequencies alone fix neither their unperturbed overtone family nor whether only cavity modes can enter the superradiant regime.

\begin{figure*}[t]
\includegraphics[width=\textwidth]{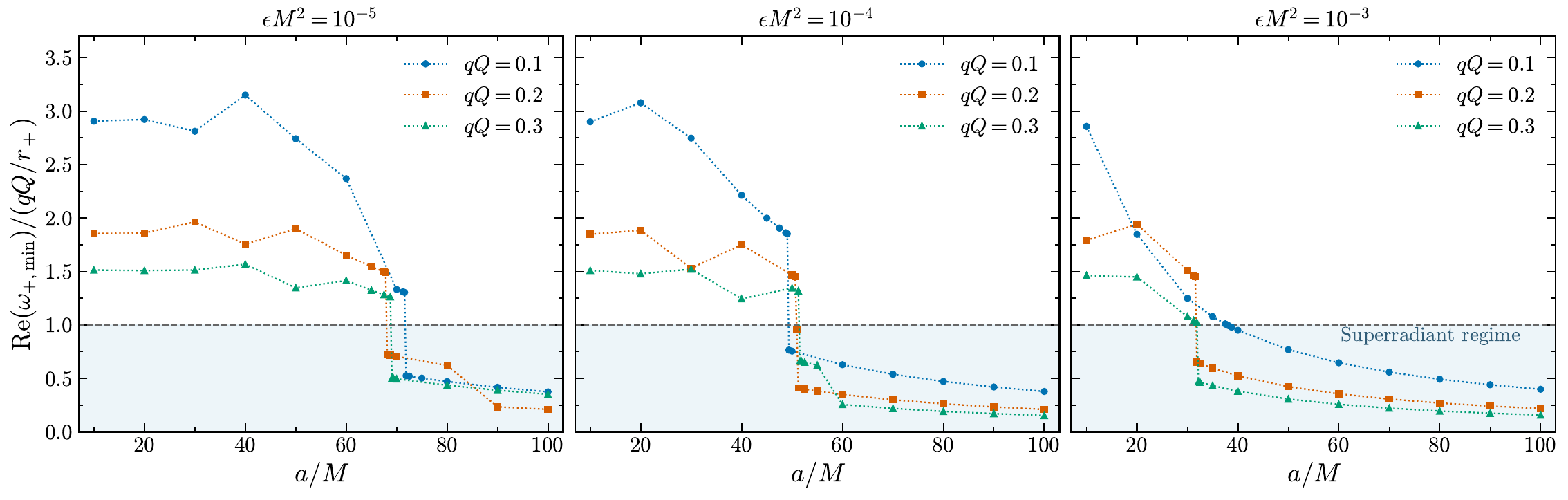}
\caption{Dependence of the selected positive-frequency RN pole on bump position and parameters. In every panel $Q/M=0.6$, $\Lambda=0$, $\ell=0$, and $w/M=1$. The columns use varying $\epsilon M^2=10^{-5},10^{-4},10^{-3}$. The differently colored curves use $qQ=0.1,0.2,0.3$. At each searched position, $\omega_{+,\min}$ minimizes $|\Imw\omega|$ among the resolved positive-frequency QNMs. The vertical coordinate $\Rew(\omega_{+,\min})/(qQ/r_+)$ measures the selected real frequency relative to the superradiant upper boundary. Values between zero and unity lie in the blue-shaded superradiant regime. The base grid is $a/M=10,20,\ldots,100$, supplemented by refinement around detected entry transitions. Dotted segments show the connection between independently ranked spectra and do not establish continuous branch identity.}
\label{fig:RNparameters}
\end{figure*}

Figure~\ref{fig:RNparameters} tests whether the entry found in Case (2) is restricted to one coupling or bump amplitude. At fixed $Q/M=0.6$, all nine displayed combinations of $qQ=0.1,0.2,0.3$ and $\epsilon M^2=10^{-5},10^{-4},10^{-3}$ have a resolved least-damped positive-frequency mode in the superradiant regime within $10\leq a/M\leq100$. Increasing the amplitude shifts the displayed entry to smaller $a/M$ for each of the three sampled couplings. These observations concern this finite parameter set and the modes resolved by the search.

Normalizing by $qQ/r_+$ places the superradiant upper boundary at unity and directly tests whether the selected real part lies in the superradiant regime. Changing $qQ$ changes both the spectrum and the normalization, so the vertical ordering of the curves is not an ordering of their unnormalized real frequencies. Abrupt changes arise when different poles exchange damping order. 

Overall, it has been shown that in the RN BH case, the monopole $\ell=0$ QNM can migrate into the superradiant regime where energy-extraction from the BH is possible, even though the initial fundamental mode lies outside the superradiant region. In fact, the initial fundamental mode is overtaken consecutively by other modes of the system as the weak bump is displaced away from the PS, thus eventually becoming superradiant but still remains damped.

\subsection{RNdS: suppressing an unstable monopole}\label{sec:dSmodes}

For RNdS we use $Q/M=0.6$, $qQ=0.2$, $\epsilon M^2=10^{-4}$, and compare $\Lambda M^2=10^{-3}$ and $10^{-4}$. The corresponding cosmological horizons are $r_c/M\simeq53.746997$ and $172.197348$, respectively. The smaller cosmological constant permits a longer areal-radius displacement while keeping the bump inside the static region. The unperturbed dS-connected \cite{Destounis:2019hca} monopole frequencies are
\begin{align}\label{eq:dsbaseline}
 M\omega_0&\simeq0.003931508+0.000298222i,
 &&\Lambda M^2=10^{-3},\nonumber\\
 M\omega_0&\simeq0.001193185+0.000097108i,
 &&\Lambda M^2=10^{-4}.
\end{align}
Both modes are growing, with instability timescales approximately $3353M$ and $10298M$.

\begin{figure*}[t]
\includegraphics[width=\textwidth]{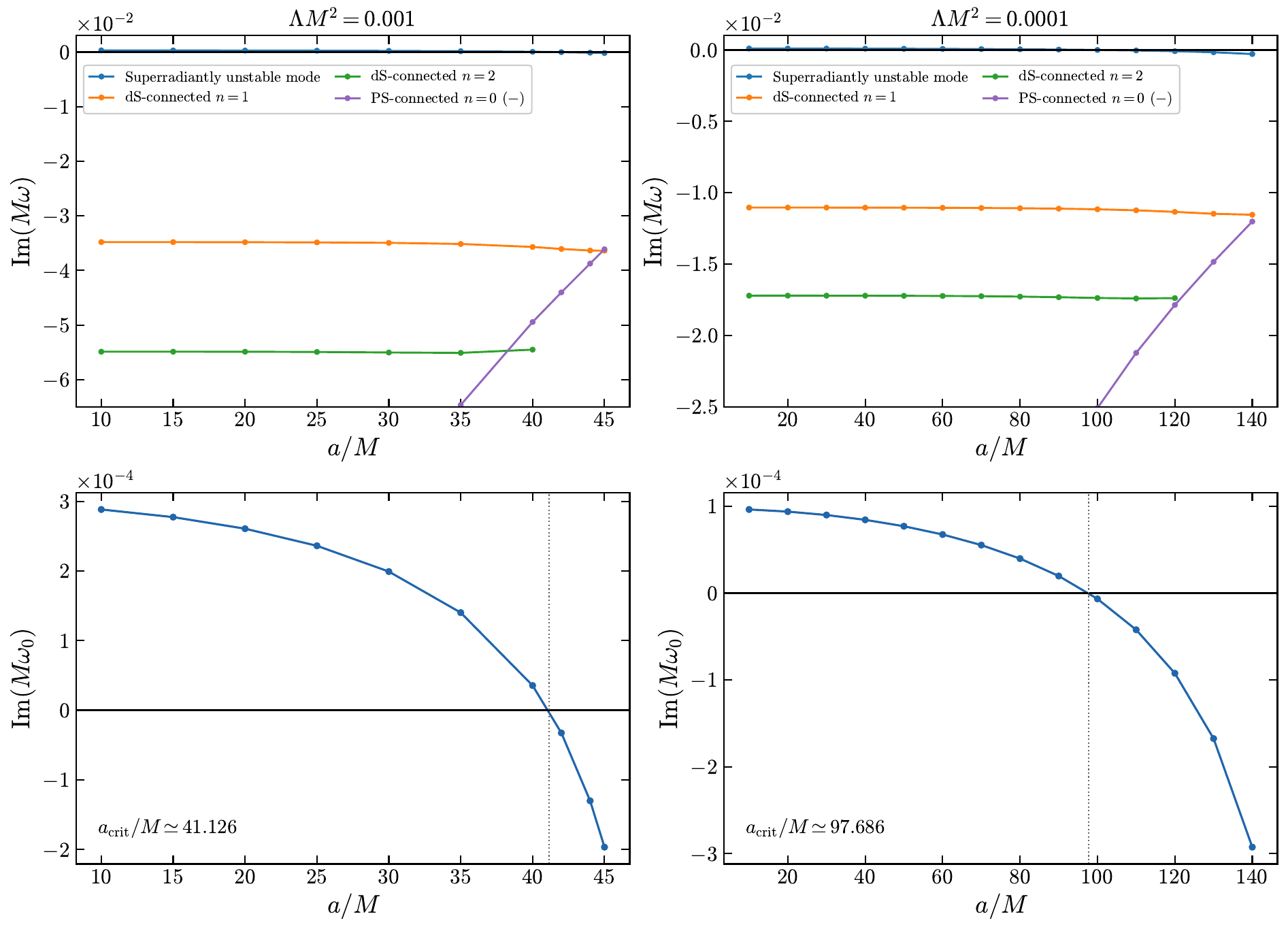}
\caption{RNdS monopole $\ell=0$ modes for $Q/M=0.6$, $qQ=0.2$, $\epsilon M^2=10^{-4}$, and bump width $M$. The columns correspond to $\Lambda M^2=10^{-3}$ and $10^{-4}$. Upper panels: Imaginary parts of the initially superradiantly unstable branch, two identified dS-connected overtones, and an identified PS-connected branch with negative real frequency, denoted PS ($-$). Colors identify the same family labels in both columns. The $n=0$ PS designation is within that branch and is not a global damping ranking. Lower panels: The imaginary part $\Imw(M\omega_0)$ behavior of the initially superradiantly unstable $\ell=0$ mode, around zero. Its sign changes at $a/M\simeq41.126$ and $97.686$, respectively. Bump locations cover $10\leq a/M\leq45$ and $10\leq a/M\leq140$. The initially unstable branch remains closest to the real axis and has the largest imaginary part among the modes retained in the searches at the sampled positions. The upper panels retain identified families only. They do not show a complete catalogue of the families of the system.}
\label{fig:DSfamilies}
\end{figure*}

Figure~\ref{fig:DSfamilies} shows how the positive bump suppresses this growth as it moves outward. The continued branch crosses the real axis at
\begin{align}\label{eq:critical}
 a_{\rm crit}/M&\simeq41.126,
 &&\Lambda M^2=10^{-3},\nonumber\\
 a_{\rm crit}/M&\simeq97.686,
 &&\Lambda M^2=10^{-4},
\end{align}
where we defined $a_{\rm crit}$ as the crossing point in areal radius of the branch. The independent root counts below test the full growing monopole spectrum at selected positions on either side; they do not map a continuous stability boundary at every intermediate position. The respective real frequencies there are $M\omega_R\simeq0.00388850$ and $0.00117980$. The critical positions depend on the bump profile, width, amplitude, and charges. They should not be considered as geometric boundaries or universal stabilization radii. Near either crossing, the growth or decay time $1/|\omega_I|$ becomes large.

Stabilization need not expel the mode from the superradiant regime. For example, at $\Lambda M^2=10^{-3}$ and $a/M=45$, the continued mode is $M\omega_0\simeq0.00386285-0.00019649i$, while $MqQ/r_c\simeq0.00372114$. At $\Lambda M^2=10^{-4}$ and $a/M=140$, it is $M\omega_0\simeq0.00114157-0.00029240i$, where its real part is below $MqQ/r_c\simeq0.00116146$. Thus, stabilization can occur while the mode remains inside the superradiant regime, and subsequent migration can also carry it below the lower boundary. Expulsion from the regime is not required at the stability transition itself. In fact, the real-frequency flux relation~\eqref{eq:neutralcurrent} requires a radiative neutral crossing with non-zero channel amplitudes to lie strictly inside that regime. The positive contribution to the potential is consistent with suppression of growth in the integrated identity~\eqref{eq:growingimag}, but the changing eigenfunction prevents a general monotonicity theorem.

The other displayed branches have been assigned to families by limiting continuations. After removing the bump and continuing to $qQ=0$, the two dS-connected overtones approach the neutral de Sitter scales associated with $-2i\sqrt{\Lambda/3}$ and $-3i\sqrt{\Lambda/3}$ as $\Lambda M^2$ decreases. These neutral limits identify the family of the charged branches, rather than approximating their frequencies at fixed $qQ=0.2$. The PS ($-$) branch is identified by removing the bump and connecting it to the negative-real-frequency RN PD branch as $\Lambda$ decreases. Other modes without established family are not discussed, but were not discarded from the underlying spectral searches. A gap or endpoint in a displayed family curve denotes the extent of the retained, identified samples. It does not imply creation or disappearance of that mode family.

No overtaking of the initially unstable branch is found among the competing modes over the sampled ranges. This contrasts with the positive-sector ordering changes in RN Case (2). Following the unstable branch alone would nevertheless be insufficient to establish modal stabilization. An independent argument-principle count gives one growing monopole at $a/M=40$ for $\Lambda M^2=10^{-3}$ and at $a/M=60$ for $\Lambda M^2=10^{-4}$, but zero at $a/M=45$ for the former and $100,140$ for the latter. These results supply numerical evidence for stabilization of the entire searched monopole problem at those configurations, rather than merely damping of one branch. 

\section{Spectral migration and amplification factors}\label{sec:resultsamp}

\subsection{RN scattering}\label{sec:RNamp}

\begin{figure*}[t]
\includegraphics[width=\textwidth]{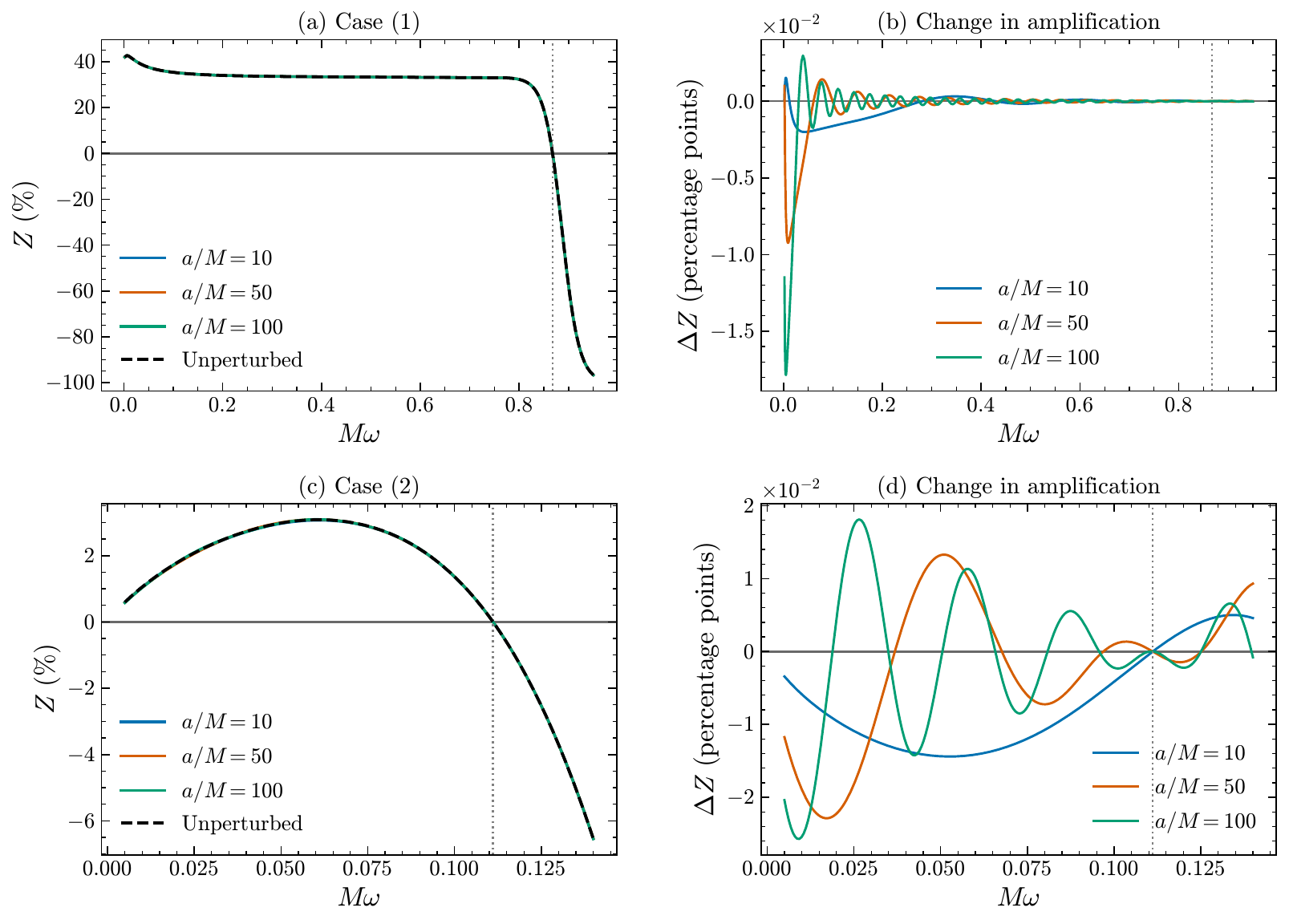}
\caption{Charged massless scalar monopole amplification in RN for Cases (1) and (2) of Table~\ref{tab:RN}, shown in the upper and lower rows. Left: Percentage gain $100Z_0$ versus $M\omega$, starting from $M\omega=0.002$ in Case (1) and $0.005$ in Case (2). Black dashed curves are unperturbed, while colored curves are perturbed by a bump with locations $a/M=10,50,100$, width $M$ and the case-dependent amplitude. The unperturbed curve is drawn over the colored curves to make their near coincidence visible Right: The corresponding changes $100\Delta Z_0$ in percentage points. Thin vertical reference lines mark $M\omega=MqQ/r_+$, i.e., the suppression of superradiance. Positive incident frequencies below this boundary are superradiant. The horizontal solid line designate zero $\Delta Z_0$. Large QNM migration in Case (2) does not produce a comparably large change or a narrow new peak in these sampled real-frequency spectra.}
\label{fig:RNamp}
\end{figure*}

Figure~\ref{fig:RNamp} compares the RN amplification spectra for the two cases in Table~\ref{tab:RN}. Each row contains the unperturbed curve and bumps at $a/M=10,50,100$. The left panels show the total percentage gain, while the right panels isolate the much smaller difference between unperturbed and perturbed RN BH potentials. The unperturbed maxima are approximately $42.67\%$ at $M\omega\simeq0.0065$ for Case (1) and $3.08\%$ at $M\omega\simeq0.061$ for Case (2). The gain is positive below the upper superradiant boundary and becomes negative above it, in agreement with Eq.~\eqref{eq:flux}. Moving the bump changes the magnitude and introduces interference structure without moving that boundary.

The environmental differences in the right panels are of order $10^{-2}$ percentage points over the plotted frequency ranges, while the continued Case-(2) frequency shifts by about $73\%$ relative to its unperturbed value. The weak scattering response in Case (2) is particularly instructive because both the original branch and the selected positive-frequency mode can enter the superradiant regime. The pole in Eq.~\eqref{eq:RNwinner} is long-lived compared with the seed mode, but is not weakly damped relative to its oscillation frequency: $\omega_R/(2|\omega_I|)\simeq0.392$. Its position therefore does not suggest the narrow, well-separated resonance associated with a high quality factor. Entry into the superradiant regime alone cannot impose such a resonance on the real-frequency response.

More generally, near a simple isolated pole a scattering amplitude can be written schematically as
\begin{equation}\label{eq:pole}
 \frac{\mathcal R}{\mathcal I}(\omega)
 \simeq B(\omega)+\frac{C}{\omega-\omega_p},
\end{equation}
where $B$ is a regular background, $C$ is the residue and $\omega_p$ the pole frequency. Even when the pole moves substantially, the measured gain also depends on $C$ and interference with $B$. We have not extracted these residues, and therefore do not attribute the small RN changes quantitatively to weak coupling or cancellation. What the direct scattering calculation establishes is that pronounced spectral migration and a modest real-frequency response coexist in the examples shown.

\subsection{RNdS scattering across stabilization}\label{sec:DSamp}

The RNdS amplification figures use $\Lambda M^2=10^{-3}$, $Q/M=0.6$, $qQ=0.2$, $\ell=0$, and $\epsilon M^2=10^{-4}$, with width $M$. Their superradiant regime is
\begin{equation}
 0.00372114<M\omega<0.11097580.
\end{equation}
The unperturbed system is modally-unstable, and its optimized scattering peak has $210.45\%$ gain at $M\omega\simeq0.00408425$. This same unstable reference is plotted in both rows of each figure.

Table~\ref{tab:RNdSresults} demonstrates the mode frequencies and optimized amplification peaks along their sequential evolution. The independent count finds exactly one growing root in each of the first two rows and none
in the last two, although the last two amplification peaks lie on opposite
sides of the $100\%$ reference.
\begin{table*}[t]
\caption{Continued monopole frequencies and optimized amplification peaks for the RNdS scattering configurations. The fixed parameters are $Q/M=0.6$, $\Lambda M^2=10^{-3}$, $qQ=0.2$, and $\ell=0$. The perturbed rows use $\epsilon M^2=10^{-4}$ and $w/M=1$. In the unperturbed row, $\epsilon=0$ and the bump position is irrelevant. The last column gives the number of growing roots found by the independent contour count of Appendix~\ref{app:counts}. The peak frequency is a real incident frequency and need not equal $\Rew\omega_0$.}
\label{tab:RNdSresults}
\begin{ruledtabular}
\begin{tabular}{lrrrrl}
Configuration & $\Rew(M\omega_0)$ & $\Imw(M\omega_0)$ & $M\omega_{\rm peak}$ & Peak gain & Growing-mode evidence\\
Unperturbed & $0.003931508$ & $+0.000298222$ & $0.00408425$ & $210.45\%$ & Count one\\
$a/M=30$ & $0.003915449$ & $+0.000199244$ & $0.00399866$ & $383.53\%$ & Count one\\
$a/M=45$ & $0.003862855$ & $-0.000196488$ & $0.00396258$ & $325.66\%$ & Count zero\\
$a/M=48$ & $0.003815589$ & $-0.000552886$ & $0.00427395$ & $71.74\%$ & Count zero
\end{tabular}
\end{ruledtabular}
\end{table*}

\begin{figure*}[t]
\includegraphics[width=\textwidth]{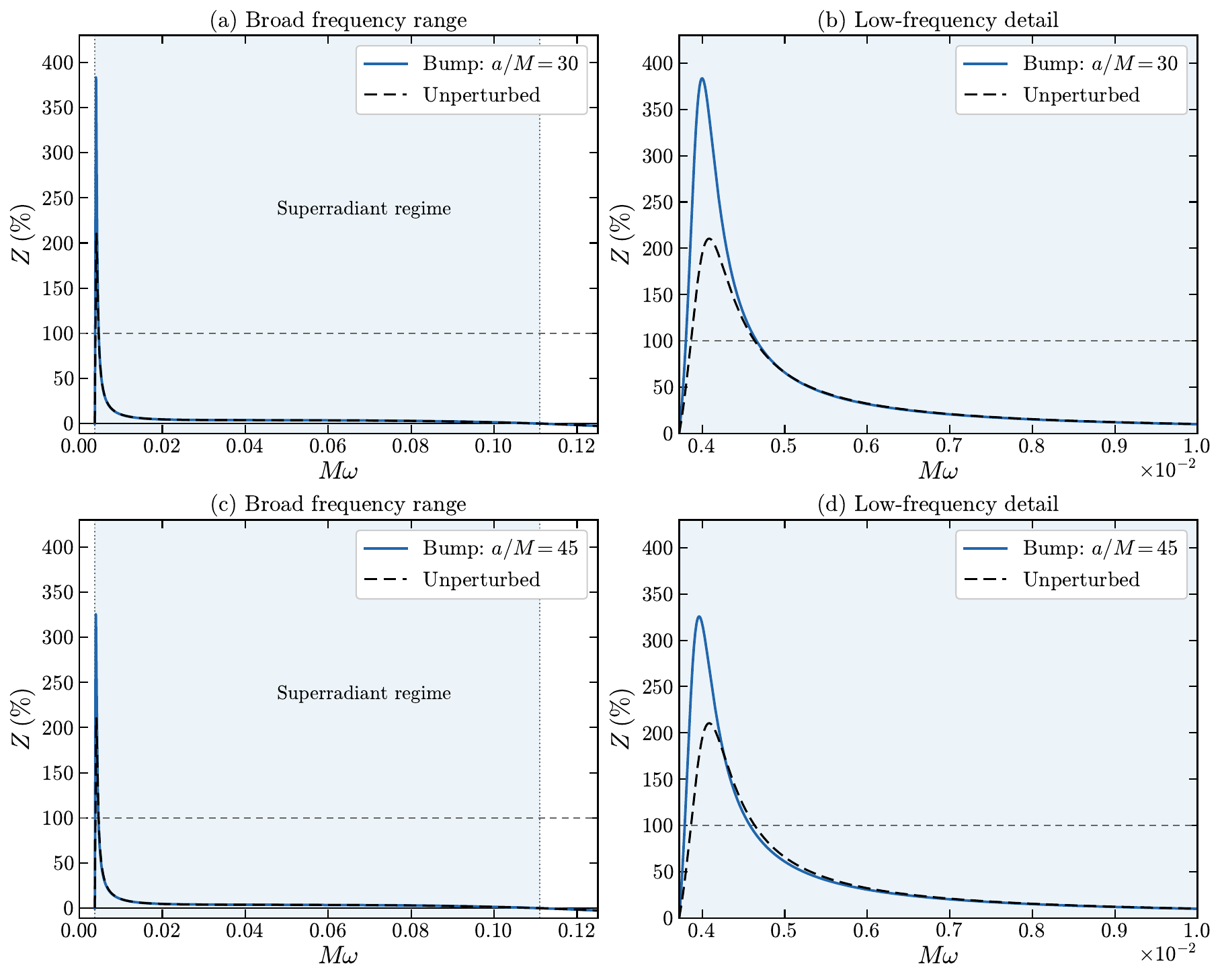}
\caption{RNdS amplification for $\Lambda M^2=10^{-3}$, $Q/M=0.6$, $qQ=0.2$, $\ell=0$, $\epsilon M^2=10^{-4}$, and $w/M=1$. The top row has $a/M=30$, below the neutral point $a_{\rm crit}/M\simeq41.126$ of the continued dS-connected branch. The bottom row has $a/M=45$, where that branch is damped and the independent growing-root count is zero. Left: broad frequency range. Right: low-frequency peak detail. Blue curves include the bump; black dashed curves show the same modally unstable unperturbed reference in both rows. The gain is $100Z_0$, with a horizontal dashed $100\%$ reference and pale-blue superradiant shading. Peak gains are approximately $383.53\%$ at $a/M=30$, $325.66\%$ at $a/M=45$, and $210.45\%$ without the bump. The independent growing-root count supports monopole stabilization at $45M$, despite an amplification peak larger than both $100\%$ and the unperturbed peak.}
\label{fig:DSamp45}
\end{figure*}

In Fig.~\ref{fig:DSamp45}, placing the bump at $a/M=30$ leaves the continued mode unstable and increases the peak gain from $210.45\%$ to $383.53\%$. At $a/M=45$, the mode becomes damped (due to spectral instability) and the growing-mode count is zero, but the peak is still $325.66\%$. This intermediate configuration is essential: stabilization does not imply suppression below the unperturbed amplification, much less below a gain of $100\%$. Table~\ref{tab:RNdSresults} also shows that the scattering peak and the real part of the pole shift by different amounts.

The proximity of a pole to the real axis helps organize this behavior. If its residue and the regular background vary smoothly, the resonant denominator in Eq.~\eqref{eq:pole} has magnitude squared
\begin{equation}\label{eq:BW}
 |\omega-\omega_p|^2=(\omega-\omega_R)^2+\omega_I^2
\end{equation}
for real incident $\omega$. It is insensitive to the sign of $\omega_I$. A near-axis pole can consequently enhance the forced response on either side of a stability transition, as in the resonant RNdS scattering studied in Ref.~\cite{Mascher:2022pku}. This local expression does not fix the exact peak frequency or height: the nonresonant background and residue can shift the maximum away from $\omega_R$. Nor do the displayed finite peaks determine a finite maximum at the exact real-axis crossing, where a coupled pole can make the stationary scattering response singular. Appendix~\ref{app:crossing} derives the conditional local scaling near a simple crossing; it is an explanation of the possible response, not a fitted law for these data.

\begin{figure*}[t]
\includegraphics[width=\textwidth]{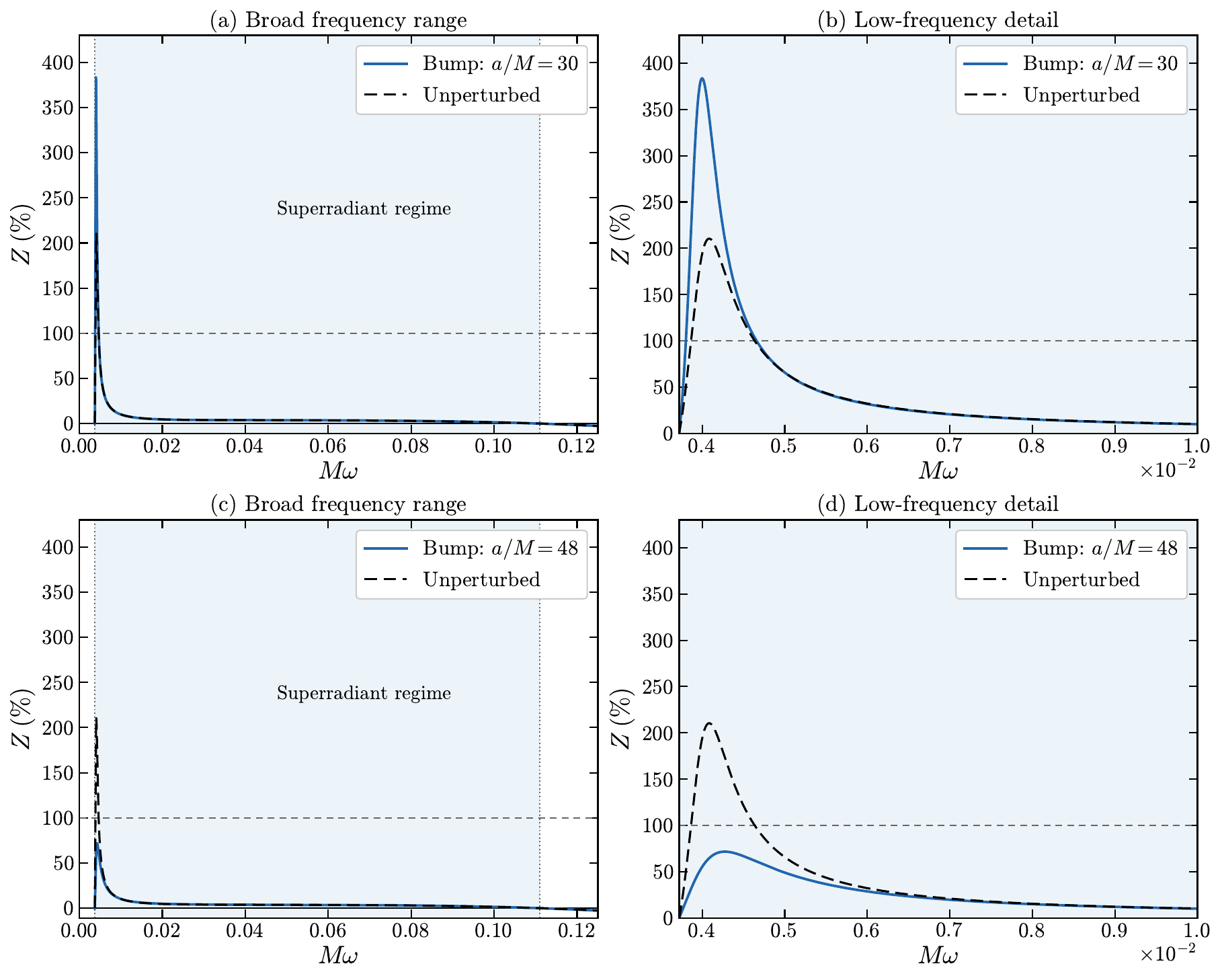}
\caption{An unstable and a numerically stabilized RNdS configuration whose peak gains lie on opposite sides of $100\%$. Parameters are $\Lambda M^2=10^{-3}$, $Q/M=0.6$, $qQ=0.2$, $\ell=0$, $\epsilon M^2=10^{-4}$, and width $M$. The upper row repeats $a/M=30$, with peak gain $383.53\%$; the lower row uses $a/M=48$, with peak gain $71.74\%$ and no growing monopole found by the independent root count. The black dashed curve in both rows is the same unstable unperturbed reference, whose peak is $210.45\%$. Left panels show the broad spectrum and right panels the low-frequency detail. The horizontal dashed line is $100\%$ gain. This pair illustrates suppression after further displacement.}
\label{fig:DSamp48}
\end{figure*}

Figure~\ref{fig:DSamp48} shows that a further displacement can substantially suppress the amplification. At $a/M=48$, the continued monopole is more strongly damped, an independent root count again finds no growing monopole, and the peak gain falls to $71.74\%$ at $M\omega\simeq0.00427395$ (Table~\ref{tab:RNdSresults}). Thus the upper and lower rows provide an unstable example with a peak above $100\%$ and a stabilized example with a peak below it. They do not establish an equivalence between those two classifications: the continued branch crosses the real axis at $a/M\simeq41.126$, whereas the optimized peak crosses $100\%$ only near $a/M\simeq47.248$ along this sequence. The intermediate $a/M=45$ configuration explicitly separates the two thresholds.

Even at $48M$, gain remains positive within the superradiant regime. Stabilization suppresses exponential growth; it does not eliminate energy extraction from incident waves. Moreover, the statements about gains above or below $100\%$ refer to the maximum over incident frequency. They do not mean that every frequency in an unstable configuration is amplified by more than $100\%$.

\subsection{Horizon separation and scope of the model}\label{sec:limitations}

The taper in Eq.~\eqref{eq:W} ensures that the horizon conditions remain exact for every plotted bump. This fact must be distinguished from a claim that the results are independent of how the tail is cut off. For the RNdS family plot, the last centers are $a/M=45$ at $\Lambda M^2=10^{-3}$ and $140$ at $10^{-4}$, leaving cosmological-horizon separations of approximately $8.747M$ and $32.197M$. The more distant center in Fig.~\ref{fig:DSamp48}, $a/M=48$, leaves $5.747M$. The corresponding untapered values at the cosmological horizon, divided by $\epsilon$, are about $1.01\times10^{-7}$, $4.32\times10^{-28}$, and $4.08\times10^{-5}$, respectively. These small tails support using a well-separated localized perturbation, but are not quantitative bounds on the spectral response.

Explicit variation of the taper width and shape has not been performed for these figures. Their numerical convergence refers to the stated tapered potential, rather than to an untapered profile or an arbitrary environmental model. The distinction becomes especially important if a bump is moved to within a few widths of the cosmological horizon, where the cutoff can reshape a substantial part of its outer tail. Such positions should not be used to infer an isolated-bump limit without an additional cutoff-sensitivity study (see Ref. \cite{Courty:2023rxk}).

There are also physical limits to the interpretation. The adopted charges and dimensionless cosmological constants are selected to resolve the spectral and scattering mechanisms; they are not a fit to an observed astrophysical environment. The varying tortoise width and local strength quantified in Sec.~\ref{sec:bump_scale} are part of this displacement experiment. A prescribed scalar-potential perturbation need not describe the propagation of other fields on a self-consistent matter distribution. Finally, neither pole migration nor stationary amplification by itself specifies a time-domain waveform, the excitation of individual modes, or a nonlinear endpoint of an instability.

\section{Conclusions}\label{sec:conclusions}

We have examined how a localized positive perturbation with small mass-scaled height affects charged massless scalar monopoles on RN and RNdS backgrounds. Tracking QNMs, searching for competeting modes, and computing real-frequency scattering responses reveal that spectral instability, modal instability, and superradiant amplification probe different aspects of the same differential equation system.

In RN, a branch displaced by the bump can enter the superradiant frequency regime while remaining damped. This conclusion survives an independent positive-frequency damping ranking: after overtaking events, the least-damped positive-frequency QNM found also enters the regime. The selected mode is not necessarily the least-damped one across both signs of real frequency, and the fixed branch labels used in the overtaking plot are not unperturbed overtone identifications.

All nine sampled RN coupling--amplitude combinations admit superradiant entry of the selected positive-frequency mode within the displayed displacement range. The RN scattering curves exhibit only modest changes for the examples studied, despite substantial pole migration in Case (2). The migrated selected positive-frequency mode is not a high-quality-factor resonance, and a real part inside the superradiant regime does not by itself enforce a narrow amplification peak. A quantitative decomposition of the scattering response would additionally require pole residues and their interference with the regular background.

In RNdS, the same type of positive perturbation can surprisingly suppress an existing superradiant monopole instability \cite{Zhu:2014sya,Konoplya:2014lha,Cardoso:2018nvb,Destounis:2019hca}. The initially growing dS-connected mode moves into the damped half-plane at parameter-dependent critical positions. Searches for other competing unstable modes and independent counts of upper-half-plane roots support stabilization at selected configurations. This is a stabilization of the linear monopole problem for the prescribed potential, but it is not a general proof of stability for all fields, all positions, or a backreacted matter geometry.

Stabilization need not reduce the amplification peak below its unperturbed value or below $100\%$. For $\Lambda M^2=10^{-3}$, $Q/M=0.6$, $qQ=0.2$, and $\epsilon M^2=10^{-4}$, the mode becomes damped near $a/M=41.126$, while the peak gain falls below $100\%$ only after $47.248$. The examples at $a=45M$ and $48M$ demonstrate these distinct stages directly. Spectral instability can, thus, suppress a growing mode while leaving substantial superradiant energy extraction available to incident waves.

The present toy model offers a tractable setting in which we can separate these effects. Further work should test sensitivity to the horizon taper and bump profile, identify the remaining competing mode branches by controlled limiting continuations, and compute residues and time-domain responses. Extending the analysis to self-consistent charged environments and other perturbing fields would establish which aspects of this stabilization mechanism persist beyond a prescribed scalar potential.

\begin{acknowledgments}
K.D. acknowledges financial support provided by FCT--Funda\c{c}\~ao para a Ci\^encia e a Tecnologia, I.P., under the Scientific Employment Stimulus -- Individual Call -- Grant No. 2023.07417.CEECIND/CP2830/CT0008. K.D. would also like to thank the Funda\c{c}\~ao para a Ci\^encia e Tecnologia (FCT), Portugal, for the financial support to the Center for Astrophysics and Gravitation (CENTRA/IST/ULisboa) through grant No. UID/PRR/00099/2025 and grant No. UID/00099/2025.
This project has received funding from the European Union's Horizon MSCA-2022 research and innovation programme ``Einstein Waves'' under grant agreement No. 101131233.
\end{acknowledgments}

\appendix
\section{Boundary factors and matching determinant}\label{app:matching}

The horizon factors make the radiative boundary conditions and their normalization explicit. Write $\kappa_+=f'(r_+)/2>0$ and $\kappa_c=-f'(r_c)/2>0$, and introduce
$x_+=(r-r_+)/M$ and $x_c=(r_c-r)/M$. In the zero-bump neighborhoods of the RNdS horizons, the ingoing and outgoing solutions have the Frobenius forms
\begin{align}\label{eq:frobenius}
 \Psi_h&=x_+^{-ik_h/(2\kappa_+)}
       \sum_{j=0}^{\infty}h_jx_+^j,\nonumber\\
 \Psi_c&=x_c^{-ik_c/(2\kappa_c)}
       \sum_{j=0}^{\infty}c_jx_c^j.
\end{align}
The exponents have the same algebraic sign because $r_*\to-\infty$ at $r_+$ but $r_*\to+\infty$ at $r_c$. The leading coefficients fix the amplitude convention. All horizon launch points lie inside the region where $\delta V=0$. At RN infinity the corresponding outgoing factor is $e^{i\omega r}(r/M)^{i(2M\omega-qQ)}$, multiplied by an inverse-radius series, consistently with Eq.~\eqref{eq:coulomb}.

Let $H$ have unit ingoing asymptotic amplitude at the event horizon and $O$ unit outgoing asymptotic amplitude at the outer boundary, with a fixed choice of the tortoise-coordinate phase. Their matching determinant is
\begin{align}\label{eq:determinant}
 D(\omega,a)&=W_*[H,O]
 =H\frac{\dd O}{\dd r_*}-\frac{\dd H}{\dd r_*}O\nonumber\\
 &=f(r)\bigl(HO'-H'O\bigr).
\end{align}
It is independent of the matching point for exact solutions of Eq.~\eqref{eq:radial}. The factor $f$ is essential when the derivatives
are evaluated in areal radius. A QNM occurs when $D=0$, since the two radiative solutions become linearly dependent.

To relate this determinant to scattering, let $I_o$ be the unit incoming outer solution and expand $H=A_{\rm in}I_o+A_{\rm out}O$. For real frequency, $W_*[I_o,O]=2ik_o$, including the asymptotic limit of the RN Coulomb waves.
Therefore, away from $k_o=0$,
\begin{align}\label{eq:incidentdet}
 D&=2ik_oA_{\rm in},\nonumber\\
 \frac{\mathcal R}{\mathcal I}
 &=\frac{A_{\rm out}}{A_{\rm in}},\qquad
 \frac{\mathcal T}{\mathcal I}=\frac{1}{A_{\rm in}}.
\end{align}
Analytic continuation supplies the connection between QNM zeros and scattering poles. Rescaling either boundary solution by a nonzero analytic factor changes the numerical value of $D$ but not its zeros. Equation~\eqref{eq:incidentdet} uses the stated unit-amplitude convention; arbitrary rescalings must be accounted for when extracting scattering amplitudes.

\section{A quadratic spectral formulation}\label{app:spectral}

In areal radius the differential equation is
\begin{equation}\label{eq:radialr}
 f^2\Psi''+ff'\Psi'
 +\bigl[(\omega-\Phi)^2-U_\ell-\delta V\bigr]\Psi=0.
\end{equation}
For RNdS, factor the radiative horizon powers by writing $\Psi=e^\chi u$, where
\begin{equation}\label{eq:spectralfactor}
 \chi=-\frac{ik_h}{2\kappa_+}\ln x_+
       -\frac{ik_c}{2\kappa_c}\ln x_c.
\end{equation}
The regular amplitude $u$ then satisfies
\begin{align}\label{eq:regularqep}
 0={}&f^2u''+(ff'+2f^2\chi')u'\nonumber\\
 &+\bigl[f^2(\chi''+\chi'^2)+ff'\chi'
        +(\omega-\Phi)^2\nonumber\\
 &\hspace{2.7cm}-U_\ell-\delta V\bigr]u.
\end{align}
Since $\chi$ is linear in $\omega$, a collocation discretization has the form
\begin{equation}\label{eq:qep}
 (A_0+\omega A_1+\omega^2A_2)\bm u=0.
\end{equation}
The horizon rows are obtained after canceling the indicial terms and removing the common vanishing factor $f$ before taking the endpoint limits. Direct substitution of $f=0$ into the unreduced expression would produce a zero row, rather than a boundary equation. Piecewise collocation resolves the bump and the taper joins, with continuity of the solution and derivative at each interface. This regularized formulation supplies candidate modes, since comparison across resolutions and independent shooting
are needed to distinguish converged poles from discretization artifacts.

\section{Growing-mode bounds and numerical root counts}\label{app:counts}

\subsection{Integral identities and the marginal mode}

For an unstable RNdS QNM, $\omega_I>0$, both radiative boundary conditions decay exponentially toward their respective horizons in $r_*$. Multiplying Eq.~\eqref{eq:radial} by $\Psi^*$ and integrating by parts gives
\begin{align}\label{eq:integrated}
 &\int_{-\infty}^{\infty}
 \left[\left|\frac{d\Psi}{d r_*}\right|^2
       +(U_0+\delta V)|\Psi|^2\right]d r_*\nonumber\\
 &\hspace{1cm}=\int_{-\infty}^{\infty}(\omega-\Phi)^2|\Psi|^2d r_*.
\end{align}
There are no boundary terms in this growing-mode case. Define
\begin{equation}\label{eq:averages}
 \langle F\rangle=
 \frac{\int F|\Psi|^2d r_*}{\int|\Psi|^2d r_*}.
\end{equation}
The imaginary part of Eq.~\eqref{eq:integrated} is
\begin{equation}\label{eq:imagidentity}
 2\omega_I\int(\omega_R-\Phi)|\Psi|^2d r_*=0.
\end{equation}
Hence $\omega_R=\langle\Phi\rangle$ and, for $qQ>0$, every growing mode obeys
\begin{equation}\label{eq:growingreal}
 \Phi_c<\omega_R<\Phi_+,\qquad
 \Phi_c=\frac{qQ}{r_c},\quad\Phi_+=\frac{qQ}{r_+}.
\end{equation}
The real part gives
\begin{align}\label{eq:growingimag}
 \omega_I^2={}&\operatorname{Var}(\Phi)-\langle U_0+\delta V\rangle\nonumber\\
 &-\frac{\int|d\Psi/d r_*|^2d r_*}
         {\int|\Psi|^2d r_*},
\end{align}
where $\operatorname{Var} (\Phi)=\langle\Phi^2\rangle-\langle\Phi\rangle^2$.
For the parameters used here, $0<f<1$ in the static region and $Mr>Q^2$. Thus
\begin{equation}\label{eq:Ubound}
 U_0=f\left[\frac{2(Mr-Q^2)}{r^4}-\frac{2\Lambda}{3}\right]
 \geq-\frac{2\Lambda}{3}.
\end{equation}
Using $\delta V\geq0$ and
$\operatorname{Var}(\Phi)\leq(\Phi_+-\Phi_c)^2/4$ gives
\begin{equation}\label{eq:bound}
 \omega_I^2\leq\frac14(\Phi_+-\Phi_c)^2+\frac{2\Lambda}{3}.
\end{equation}
This bound constrains every growing monopole independently of its family. At $Q/M=0.6$ and $qQ=0.2$, the upper bounds on $M\omega_I$ are approximately $0.0595194$ and $0.0555712$ for $\Lambda M^2=10^{-3}$ and $10^{-4}$,
respectively. These are bounds on possible growth rates, not measured mode frequencies.

The term $-\langle\delta V\rangle$ in Eq.~\eqref{eq:growingimag} is consistent with suppression of growth by a positive bump. It is not a monotonicity theorem in $a$ or $\epsilon$, since the mode profile, the variance, and the gradient term also change. The direction and magnitude of the spectral response must therefore be determined from the boundary-value problem.

Equations~\eqref{eq:integrated}--\eqref{eq:growingimag} must not be applied
directly to marginal or damped QNMs. At $\omega_I=0$ the horizon amplitudes generally do not decay and the integrals are not finite. Instead, for a real-frequency QNM at $\omega=\omega_{\rm crit}$, set $\mathcal I=0$ in the conserved-current relation to obtain
\begin{equation}\label{eq:neutralcurrent}
 (\omega_{\rm crit}-\Phi_c)|\mathcal R|^2
 =-(\omega_{\rm crit}-\Phi_+)|\mathcal T|^2.
\end{equation}
For nonzero radiative amplitudes, away from the degenerate channel thresholds, the two frequency factors have opposite signs. A real-axis radiative crossing consequently lies within $\Phi_c<\omega_{\rm crit}<\Phi_+$. This conclusion does not require extending the square-integrable growing-mode identity to a neutral mode.

\subsection{Counting growing roots}

For a counterclockwise contour $\mathcal C$ on which $D$ has neither zeros nor poles, the argument principle states
\begin{equation}
\label{eq:argumentprinciple}
 N_Z-N_P=
 \frac{1}{2\pi i}\oint_{\mathcal C}
 \frac{\partial_\omega D}{D}\,d\omega
 =\frac{\Delta_{\mathcal C}\arg D}{2\pi}.
\end{equation}
The winding is a zero count only after poles of the chosen determinant have been excluded or included in $N_P$. With fixed nonzero leading horizon coefficients, the exceptional frequencies of the RNdS Frobenius recurrences occur at $\omega=\Phi_j-in\kappa_j$, with $j\in\{+,c\}$ and $n=1,2,\ldots$, below the real axis. A further division by a frequency-dependent solution value at a finite point can introduce poles and must be accounted for. Analytic, zero-free propagation factors leave the winding unchanged. Positive real
rescalings at sampled contour points preserve phase too, although the rescaled samples need not define an analytic function.

The growing-root check evaluates the horizon-normalized determinant on a rectangle enclosing the region allowed by Eqs.~\eqref{eq:growingreal} and~\eqref{eq:bound}. Its real-axis lower edge is used away from the critical configurations at which a root lies on that edge. The upper and side edges must lie beyond the analytic bounds, rather than merely
outside a chosen damped-mode search window. The recorded counts agree between two successively refined contour samplings and are summarized in Table~\ref{tab:counts}. They are numerical counts, not interval-arithmetic proofs.

\begin{table}[t]
\caption{Recorded numbers $N_+$ of growing monopoles for $Q/M=0.6$, $qQ=0.2$, $\epsilon M^2=10^{-4}$, $w=M$, and the taper in
Eq.~\eqref{eq:W}; the first row is the unperturbed system ($\epsilon=0$). Only the listed configurations have these independent
counts; the table does not assert stability at all intermediate positions.}
\label{tab:counts}
\begin{ruledtabular}
\begin{tabular}{ccc}
$\Lambda M^2$ & $a/M$ & $N_+$ \\
\hline
$10^{-3}$ & -- & $1$ \\
$10^{-3}$ & $30$ & $1$ \\
$10^{-3}$ & $40$ & $1$ \\
$10^{-3}$ & $45$ & $0$ \\
$10^{-3}$ & $48$ & $0$ \\
$10^{-4}$ & $60$ & $1$ \\
$10^{-4}$ & $100$ & $0$ \\
$10^{-4}$ & $140$ & $0$
\end{tabular}
\end{ruledtabular}
\end{table}

The count summary does not quantify the phase-resolution margin of each contour. Reproducible exclusion of missed windings requires the actual vertices, mesh sizes, maximum resolved phase increments, and a
normalization-consistent measure of proximity to zero, together with the frequency-convergence records. These diagnostics are not specified by agreement of two integer counts alone.

Finite-resolution spectral searches locate damped competitors and check the selected ordering, but do not establish completeness of all damped poles. An unvalidated candidate is not assigned a physical family merely
because its residual is small. The stabilization evidence uses the separate
growing-root counts at the listed configurations. 

\section{Local behavior near a simple neutral point}\label{app:crossing}

The connection between a growth-rate crossing and a large scattering response can be made explicit under a simple-pole assumption. Suppose $D(\omega_{\rm crit},a_{\rm crit})=0$, with $\omega_{\rm crit}$ real and
$\partial_\omega D\ne0$. Implicit differentiation gives
\begin{align}
\label{eq:crossingslope}
 \omega_p(a)&=\omega_{\rm crit}+s(a-a_{\rm crit})
             +O\bigl((a-a_{\rm crit})^2\bigr),\nonumber\\
 s&=-\left.\frac{\partial_aD}{\partial_\omega D}
        \right|_{(\omega_{\rm crit},a_{\rm crit})}.
\end{align}
If $\Imw s\ne0$, the growth or decay time diverges at leading order as $1/(|\Imw s|\,|a-a_{\rm crit}|)$. For an isolated pole with nonzero scattering residue $C_{\rm crit}$, negligible leading background contribution, and a peak remaining inside the allowed incident-frequency domain, Eq.~\eqref{eq:pole} yields
\begin{equation}
\label{eq:peakscaling}
 1+Z_{\max}\sim
 \frac{|C_{\rm crit}|^2}{(\Imw s)^2(a-a_{\rm crit})^2}.
\end{equation}
This conditional local estimate explains why strong amplification can persist on either side of a change in the sign of growth. It is not a measured scaling law for the present data and does not identify the $100\%$ crossing with the neutral point. Residue variation, interference, nearby modes, and channel thresholds require the full scattering response.

\bibliography{biblio}

\end{document}